\documentclass[prd,twocolumn,aps,showpacs,tightenlines,superscriptaddress,amsmath,amssymb,amsfonts, nofootinbib]{revtex4-2}
\usepackage{amssymb,amsmath,amsthm,amsbsy,epsfig,color,graphicx,times,physics}
\usepackage{subcaption}
\usepackage[ansinew]{inputenc}
\usepackage[english]{babel}
\usepackage{color}
\usepackage{braket}
\usepackage{tabularx}
\usepackage{array}
\usepackage{slashed}
\usepackage{tikz}
\usetikzlibrary{arrows.meta, positioning}
\usepackage{lmodern}
\usepackage{float}

\begin{document}

\title{Single arm interferometry to probe  the scalar field dark matter}

\author{A. Capolupo}
\email{capolupo@sa.infn.it}

\author{G.Pisacane}
\email{gpisacane@unisa.it}
\affiliation{Dipartimento di Fisica ``E.R. Caianiello'' Universit\'a di Salerno,
and INFN -- Gruppo Collegato di Salerno, Via Giovanni Paolo II, 132,
84084 Fisciano (SA), Italy}

\author{A. Quaranta}
\email{aniello.quaranta@unicam.it}
\affiliation{School of Science and Technology, University of Camerino, Via Madonna delle Carceri, Camerino, 62032, Italy}

\author{R. Serao}
\email{rserao@unisa.it}
\affiliation{Dipartimento di Fisica ``E.R. Caianiello'' Universit\'a di Salerno,
and INFN -- Gruppo Collegato di Salerno, Via Giovanni Paolo II, 132,
84084 Fisciano (SA), Italy}

\begin{abstract}

We analyze the interaction of photons with a scalar dark matter field $\phi$ and we propose to use  a single arm interferometer to reveal this interaction and constrain the parameters of the scalar dark matter model. By considering a beam of coherent light and two spatially separated squeezing operations, we show that the interaction of photons with scalar dark matter leads to an observable deviation in the outgoing light state, with respect to free evolution. Therefore the single arm interferometer may yield a novel revelation method for scalar dark matter.

\end{abstract}

\maketitle

 \section{Introduction}
  
  The nature of the dark sector of the universe is one of the main puzzles in modern cosmology.  Dark energy is held responsible for  the observed accelerated expansion of the universe, while dark matter is required to explain the gravitational binding of galaxies and galaxy clusters, as well as the rotation curve of spiral galaxies \cite{DM1,DM2,DM3,DM4,DM5,DM6,DM7,DM8}.
 Prominent dark matter candidates are axions and axion-like particles \cite{AX1,AX2,AX3,AX4,AX5,AX6,AX7,AX8,AX9,AX10,AX11,AXE1,AXE2,AXE3,AXE4,AXE5,AXE6,AXE7,AXE8,AXE9,AXE10,AXE11,AXE12,AXE13,AXE14}, supersymmetric particles \cite{susy1,susy2,susy3,susy4,susy5,susy6,susy7,susy8,susy9,susy10,susy11, susy12,susy13,susy14,susy15,susy16,susy17}, ultra-light scalar particles, known as scalar dark matter (SDM) \cite{DM10,SDM1,SDM2,SDM3,SDM4,SDM5,SDM6,SDM7,SDM8,SDM9,SDM11}, and fuzzy cold dark matter  \cite{Fuzzy}. Condensates from quantum field theory have also been suggested as possible contributions to the dark matter \cite{BSM1,BSM2,BSM3,BSM4,BSM5,BSM6,BSM7}.
  In this work we focus our attention on SDM. Several techniques have been proposed to test the SDM hypothesis. These include the use of highly sensitive optical cavities \cite{OP1}, gravitational wave detectors  \cite{GW1,GW2,GW3}, and analyses of astrophysical phenomena that could provide indirect evidence for its existence \cite{AE1,AE2}.
Constraints on  SDM  have been obtained in various experiments such as DAMA   \cite{DAMA1,DAMA2} and   precision   atomic and molecular spectroscopy \cite{ATOM1,ATOM2,ATOM3,ATOM4,ATOM5}.

Here, we present a novel approach to test the SDM model by means of single arm laser interferometry  \cite{INT1}.
The method is aimed at probing the phase shift and the intensity variation, experienced by a laser beam due to the interaction with the SDM. By considering laser light passing through two spatially separated squeezing operations, we show that the interaction between photons and scalar dark matter causes a detectable alteration in the outgoing laser state compared to its free evolution. This interaction cannot be revealed in standard interferometry experiments, since it affects the evolution of photons in the two arms equally and, therefore, no effect can be detected by phase difference analyses. In the following, we provide  a discussion of the interaction between SDM and photons, then we introduce  the interferometric setup, and compute the phase shift and the intensity variation due to SDM-photon interaction. By using single arm interferometers with an effective
long baseline, similar to the one of the gravitational wave detector arm cavities, it is possible to verify the existence of the coupling between SDM and photons and, consequently, test the presence of scalar dark matter.

In Sec. II, we introduce the interaction between scalar dark matter and photons and examine its impact on the evolution of laser light. In Sec. III, we demonstrate how this interaction can be probed by means of a coherent light beam and two spatially separated squeezing operations within a single-arm interferometric setup. Numerical results are presented in Sec. IV, followed by a discussion of the implications and our conclusions in Sec. V.

 \section{Scalar dark matter and photon interaction}
 
The evolution of the scalar field $\phi$ is described by the Klein-Gordon equation: $\left(\Box + m_{\phi}^2\right) \phi = 0$, where $m_{\phi}$ represents the mass of the field.
A key aspect of SDM's phenomenology is its possible interaction with the photon field. This interaction can be described by an effective coupling term in the Lagrangian density \footnote{The total interaction Lagrangian when also the electron field is present is given by: $\mathcal{L}_{\text{int}} = \frac{1}{4}  \frac{\phi}{\Lambda_\gamma} F_{\mu \nu} F^{\mu \nu}-\frac{\phi}{\Lambda_e}\Bar{\psi_e}\psi_e$. Since we consider the propagation of the photon in vacuum, the last term is irrelevant, at least to lowest order.} $\mathcal{L}_{\text{int}} =\frac{1}{4} \frac{\phi}{\Lambda_\gamma} F_{\mu \nu} F^{\mu \nu}$ \cite{RING,ED7}, where $F_{\mu \nu}$ is the electromagnetic field tensor and $\Lambda_\gamma$ parametrizes the interaction strength \footnote{ We point out that, in classical mechanics on electromagnetic field configurations for which $F^2=0$ neither the photon nor the scalar field are affected by this interaction. On the contrary in quantum mechanics both the Hamiltonian and $F^2$ are operators that don’t have a definite value, except when evaluated on specific states. This in turn means that the relation holding classically for plane wave $F^2=0$ does not hold as an operator identity $\hat{F}^2\neq0$, as can be immediately verified by inserting the photon field expansion within the photon lagrangian. It follows that quantum mechanically both the photon and the scalar fields are affected by interaction term $\mathcal{L}_{\text{int}} = \frac{1}{4}  \frac{\phi}{\Lambda_\gamma} F_{\mu \nu} F^{\mu \nu}$.}.
The interaction term affects the propagation of photons through regions permeated by SDM. The field $\phi$ is assumed to have an extremely small mass ($\ll 1 eV$) \cite{DM10}; then its Compton wavelength $\lambda_c = \frac{h}{m c} $ is very large, and the field $\phi$ can be approximated as coherently oscillating   classical field: $\phi(t) \approx \phi_0 \cos(\omega t + \tilde{\theta})$. Here, $\tilde{\theta}$ is a random phase factor,  $\phi_0 = \frac{1}{m_\phi} \sqrt{2 \rho_{\text{local}}}  $ is the amplitude of the field,  with $\rho_{\text{local}}$ local dark matter density,  $m_\phi$ is the mass of the scalar particle and $\omega$ is the oscillation frequency, which is of the order of the mass of the SDM, $\omega\simeq m_\phi$ \cite{GW2,ED2,AE1}. We consider  natural units.\\

We impose the radiation gauge, where the electromagnetic potential $A^\mu=\left( A^0,\vec{A} \right)$  satisfies the condition $\vec{\nabla}\cdot\vec{A}=0$. Then, the Hamiltonian corresponding to the above Lagrangian density $\mathcal{L}_{\text{int}}$ is\footnote{Notice that the interaction Lagrangian, $\mathcal{L}_{int}$, can be written in terms of the free electromagnetic Lagrangian, $\mathcal{L}_0=-\frac{1}{4}F_{\mu \nu}F^{\mu \nu}=\frac{1}{2}(\vec{E}^2-\vec{B}^2)$, as $\mathcal{L}_{\text{int}}=-\frac{\phi(t)}{\Lambda_\gamma}\mathcal{L}_0$. As a consequence, $\mathcal{H}_{\text{int}}=-\frac{\phi(t)}{\Lambda_\gamma}\mathcal{H}_0$, where $\mathcal{H}_0$ is the free electromagnetic Hamiltonian. Since $\mathcal{H}_0=\frac{1}{2}(\vec{E}^2+\vec{B}^2)$, then $\mathcal{H}_{\text{int}}\neq 0$  even when $F^2=0$. This consideration is independent of the classical or quantum mechanical nature of the electromagnetic field. However, in our computation we consider the quantized Hamiltonian which, a fortiori, is different from zero (see also Footnote 2). Since in the following we will analyze the quantum dynamics of the photons, it is necessary to employ the quantum mechanical Hamiltonian operator (See eq. \eqref{Ham}).  }:
\begin{align*}
	H_{\text{int}}(t)
	= -\,\frac{\phi(t)}{2\;\Lambda_\gamma}
	\!\int d^3x\, \left[ \left( \dot{\vec{A}}\right)^2+\left(\vec{\nabla} \cross \vec{A}\right)^2\right]
\end{align*}
Upon quantization, the classical fields are replaced by the corresponding field operators, and the interaction Hamiltonian becomes an operator written in terms of the photon creation and annihilation operators. Normal ordering is then applied to eliminate the vacuum expectation value of the field energy. The resulting interaction Hamiltonian operator reads
\begin{eqnarray} \label{Ham}
	\hat{H}_{\text{int}}(t)
	= -\,\frac{\phi(t)}{\Lambda_\gamma}
	\int d^3k \sum_{\lambda}
	|\vec{k}|\,
	\hat{N}_{\vec{k}\lambda},
\end{eqnarray}
where $\hat{N}_{\vec{k} \lambda} = \hat{a}_{\vec{k}\lambda}^\dagger \hat{a}_{\vec{k}\lambda}$ is the  number operator of photons with momentum $\vec{k}$ and polarization $\lambda$ and $\hat{a}_{\vec{k}\lambda}$ is the annihilator.
States with a definite number of photons are eigenstates of the Hamiltonian \eqref{Ham}. Consequently, if the initial photon state is $\ket{n_{\vec{k}\lambda}}=\frac{1}{\sqrt{n!}}\left(\hat{a}_{\vec{k}\lambda}^\dagger \right)^n\ket{0}$, the evolved state at time $t$, $\ket{n_{\vec{k}\lambda}(t)}$, in interaction picture is:
\begin{equation}\label{num}
    \ket{n_{\vec{k}\lambda}(t)} =  e^{-i\int_{0}^{t}\hat{H}_{\text{int}}(t') \, dt'} \ket{n_{\vec{k}\lambda}} = e^{i \hat{N}_{\vec{k}\lambda} \, \delta_k(t)} \ket{n_{\vec{k}\lambda}},
\end{equation}
with  phase shift $\delta_k(t)$ given by:
\begin{equation}\label{phase}
    \delta_k(t) =\frac{\left |\vec{k}\right|}{\Lambda_\gamma}\int_{0}^{t}\phi(t') dt'= \frac{\left |\vec{k}\right|\phi_0 }{\Lambda_\gamma \; \omega} \sin(\omega t)\;.
\end{equation}
Note that the phase shift is independent on the polarization $\lambda$.
In the following we shall consider polarized (fixed $\lambda$) and monochromatic (fixed $k$) light and we will omit the indices $k,\lambda$. To study the evolution of laser light, we analyze a coherent state $\ket{\alpha} = e^{-\frac{|\alpha|^2}{2}} \sum_{n=0}^{\infty} \frac{\alpha^n}{\sqrt{n!}} \ket{n}$. For coherent states,  one has
$e^{i \hat{N} \delta}\ket{\alpha} = e^{-\frac{|\alpha|^2}{2}} \sum_{n=0}^{\infty} \frac{\alpha^n e^{i n\delta}}{\sqrt{n!}} \ket{n}
= \ket{\alpha e^{i\delta}}$, i.e.
  \(e^{i\hat{N}\delta}\) rotates the complex amplitude \(\alpha\) by an angle \(\delta\) in the complex plane.

\section{Phase measurement scheme}

\begin{figure}[H]
\centering
\includegraphics[width=1\linewidth]{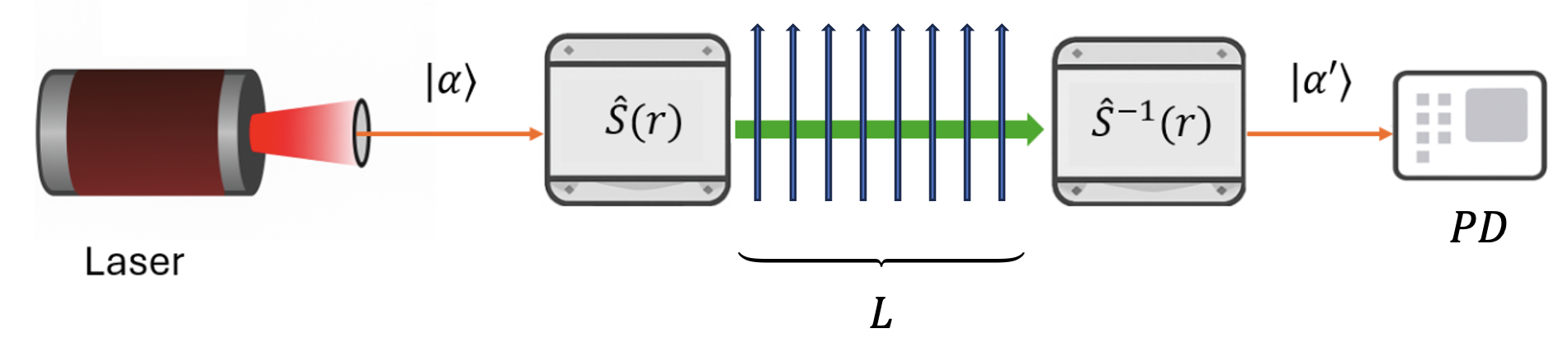}
\caption{(Color online). Schematic depiction of the single-arm interferometer. A coherent laser beam, represented by the quantum state $ \ket{\alpha} $, is emitted from the laser source. The beam first undergoes a squeezing operation $ \hat{S}(r) $. This is followed by an inverse squeezing (antisqueezing) operation $ \hat{S}^{-1}(r) $, which give the state $ \ket{\alpha'} $ which is probed by a photodetector ($PD$).}
    \label{schema}
\end{figure}

The phase shift in eq. \eqref{phase} cannot be observed using a double-arm interferometer like the Michelson-Morley type, since the interaction Lagrangian $\mathcal{L}_{\text{int}}$ influences both arms in the same way, leading to the cancellation of the phase shifts \cite{INT1}. However, a single-arm interferometer like the one described in refs. \cite{INT2, INT3, INT4}, can be utilized for our purposes. The single arm interferometer can be schematized as follows. First a squeezing operation $\hat{S}$ is performed on the laser beam. The beam then evolves along a path of length $L$, in vacuum, subject to the interaction with SDM. Later an anti-squeezing operation $\hat{S}^{-1}$ is applied and the laser light is detected, see Fig.\ref{schema}.
If the beam does not undergo interaction with scalar fields, the initial state of the photons coincides, up to a global phase, with the final state since the beam is subjected to the action of $\hat{S}$ and $\hat{S}^{-1}$. If, however, in the path $L$ the photons interact with the scalar field, the final state will differ from the initial one. Thus, an eventual revelation of the difference between the initial and final states could signal the presence of SDM interacting with photons. After the squeezing operation, the squeezed state that we employe are bright, that is carrying a large average number of photons, and then distinct from the squeezed vacuum. For such states standard read/out techniques, RF modulation/demodulation or DC readout techniques can be used to stabilize the phase, as reported in \cite{Doli?ska,mc,Simon}. We point out that the coherent control loop will not be affected by SDM.
In particular, we consider an initial coherent beam of light $\ket{\alpha}$; it satisfies the condition,  $\hat{a}\ket{\alpha}=\alpha\ket{\alpha}$, where $\alpha=\left|\alpha \right| e^{i\theta}$, and the average photon number is given by $\bra{\alpha}\hat{a}^\dagger \hat{a}\ket{\alpha}=|\alpha|^2=N$.
According to eq.\eqref{num}, the state $\ket{\alpha}$  at time $t$, is  $e^{i \hat{N} \delta(t)}\ket{\alpha}$, then, at any time $t$, the average photon number is constant, indeed
 $
   \bra{\alpha(t)} \hat{a}^\dagger \hat{a}\ket{\alpha(t)}=\bra{\alpha} e^{-i \hat{N} \delta(t)}\hat{a}^\dagger \hat{a}e^{i \hat{N} \delta(t)}\ket{\alpha}=|\alpha|^2.
$
Therefore, the presence of the interaction in eq.\eqref{Ham} cannot be investigated by analyzing the simple  time evolution of the number operator.
In the  system of Fig. \ref{schema}, the squeezing operator $\hat{S}$ is
$
    \hat{S}(r) = \exp\left[\frac{r}{2} \left(\hat{a}^{\dagger 2} - \hat{a}^2 \right)\right],
$
with $ r $   the squeezing parameter. The squeezed state, obtained by the application of $\hat{S}$,  is then $\ket{\Phi(r)}=\hat{S}(r)\ket{\alpha}$.
With reference to the device sketched in Fig. \ref{schema},  the emergent state $|\alpha'\rangle$
in  absence of interaction, would be: $|\alpha'\rangle_{\delta = 0} = \hat{S}^{-1}(r) \hat{S}(r) |\alpha\rangle = |\alpha\rangle$. On the contrary,  in presence of the interaction \eqref{Ham}, $|\alpha'\rangle$  experiences a phase shift:
$
|\alpha'\rangle = \hat{S}^{-1}(r) e^{i \delta \hat{a}^\dagger \hat{a}} \hat{S}(r) |\alpha\rangle,
$
where $\delta$ is the phase of eq. \eqref{eq:phase_shift}.  From the squeezing transforations \footnote{It is easy to check that, if there is no mixing between annihilation and creation operators, the interaction in Eq.~\eqref{Ham} does not produce any observable effects. A squeezing-like operation, such as the Bogoliubov transformation in Eq.~\eqref{4}, is necessary to detect the effects of the interaction in Eq.~\eqref{Ham} within a single-arm interferometer.
}
\begin{equation}\label{4}
\begin{split}
&\hat{S}^{-1}(r) \hat{a}\hat{S}(r) = \hat{a} \cosh r + \hat{a}^\dagger \sinh r,\\
&\hat{S}^{-1}(r) \hat{a}^\dagger \hat{S}(r) = \hat{a}^\dagger \cosh r + \hat{a}\sinh r,\\
\end{split}
\end{equation}
Here  $\cosh r$ and  $\sinh r$ are Bogoliubov's coeffients and the number of photons emerging after the application of the squeezing operators $\hat{S}$ and
 $\hat{S}^{-1}$ will be
\begin{eqnarray}\label{Numm}
\nonumber N_{out}(t) & =& \bra{\alpha'} \hat{a}^\dagger \hat{a} \ket{\alpha'}\\
\nonumber & =& |\alpha|^2\left[1 - \sinh(2r)\sin(2\theta)\sin(2\delta(t)) + \right.\\
\nonumber &+& \left.\sinh(4r)\cos(2\theta)\sin^2\delta(t) + 2\sinh^2(2r)\sin^2\delta(t)\right] \\
&+& \sinh^2(2r)\sin^2\delta(t),
\end{eqnarray}
in the following we set $\theta=0$.
Notice that for $\delta=0$, one has  $N_{out}=N_{in}=|\alpha|^2$.
Therefore,  $N_{out} \neq N_{in},$ provides an indirect evidence of a non-zero phase shift induced by the SDM-photon interaction.

	
Introducing the relations:
$\phi_0 = \frac{1}{m_\phi} \sqrt{2 \rho_{\text{local}}}, $ and $ \omega = m_\phi,$  in eq.\eqref{phase}, and being   (in natural units)   $t = L$, the phase shift \eqref{phase} induced by the scalar field becames:
\begin{equation}\label{eq:phase_shift}
    \delta(L) = \frac{ g_{\gamma}\sqrt{2 \rho_{\text{local}}}}{ \; m_\phi^2} |\vec{k}|  \sin\left(m_\phi  L\right) ,
\end{equation}
where $g_{\gamma}=1/\Lambda_\gamma$ and $\left|\vec{k}\right| = 2\pi / \lambda$ is the photon wavenumber (with wavelength $\lambda$).
Eq.(\ref{eq:phase_shift}) shows that the phase shift depends on the local dark matter density $\rho_{\text{local}}$, the photon momentum $\left|\vec{k}\right|$, the scalar field mass $m_\phi$, and the coupling constant $g_\gamma$.

We emphasize that, to observe the interaction between SDM and photons, it is necessary to have a single-arm interferometer with a long effective optical path, as can be achieved in experimental setups such as LIGO, VIRGO and PVLAS. LIGO is a two-arm Fabry-Perot-Michelson interferometer that precisely measures the differential arm length. For our purposes, the current configuration cannot be used. One possible use of devices like LIGO and VIRGO would consist of using only a single arm of these interferometers, applying squeezing and antisqueezing operators at the beginning and end of the effective optical path, and using a photodetector to count the output  photons.  In the case of PLVAS, however, the device is represented by a Fabry-Perot cavity, realizing a long effective optical path, and the photodetectror is already implemented in the setup. Therefore, it would only be necessary to apply the squeezer and antisqueezer operators, removing the magnetic field used in PVLAS and not necessary for our studies. In the following we consider values of the parameters compatible with the LIGO interferometer.

\section{Numerical analysis} 

\begin{figure}[H]
	\centering
	\includegraphics[width=1\linewidth]{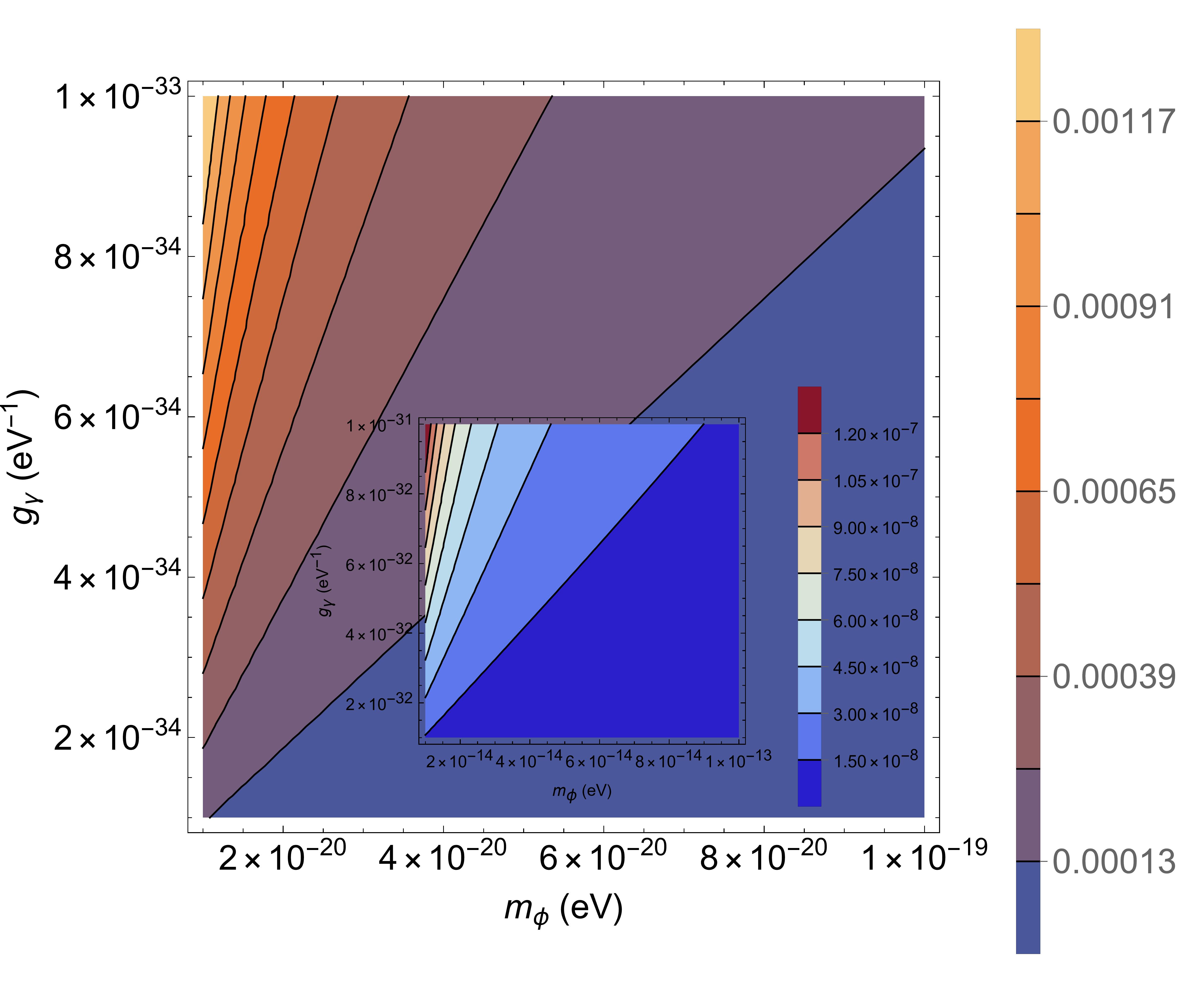}
	\caption{(Color online). Main picture: plot of $\delta(L)$ as function of $g_\gamma\in [10^{-34}, 10^{-33}] \, \mathrm{eV^{-1}}$ and $m_\phi\in [10^{-20}, 10^{-19}] \, \mathrm{eV}$.
 In the inset is plotted $\delta(L)$ as function of $g_\gamma\in [10^{-31}, 10^{-30}] \, \mathrm{eV^{-1}}$ and $m_\phi\in [10^{-13}, 10^{-12}] \, \mathrm{eV}$. We consider the arm length $L= 950\;\mathrm{km}$ and the photon wavelength $\lambda=1064\cdot 10^{-9}\;\mathrm{m}$.}
	\label{fig2}
\end{figure}

\begin{figure}[H]
	\centering
	\includegraphics[width=1\linewidth]{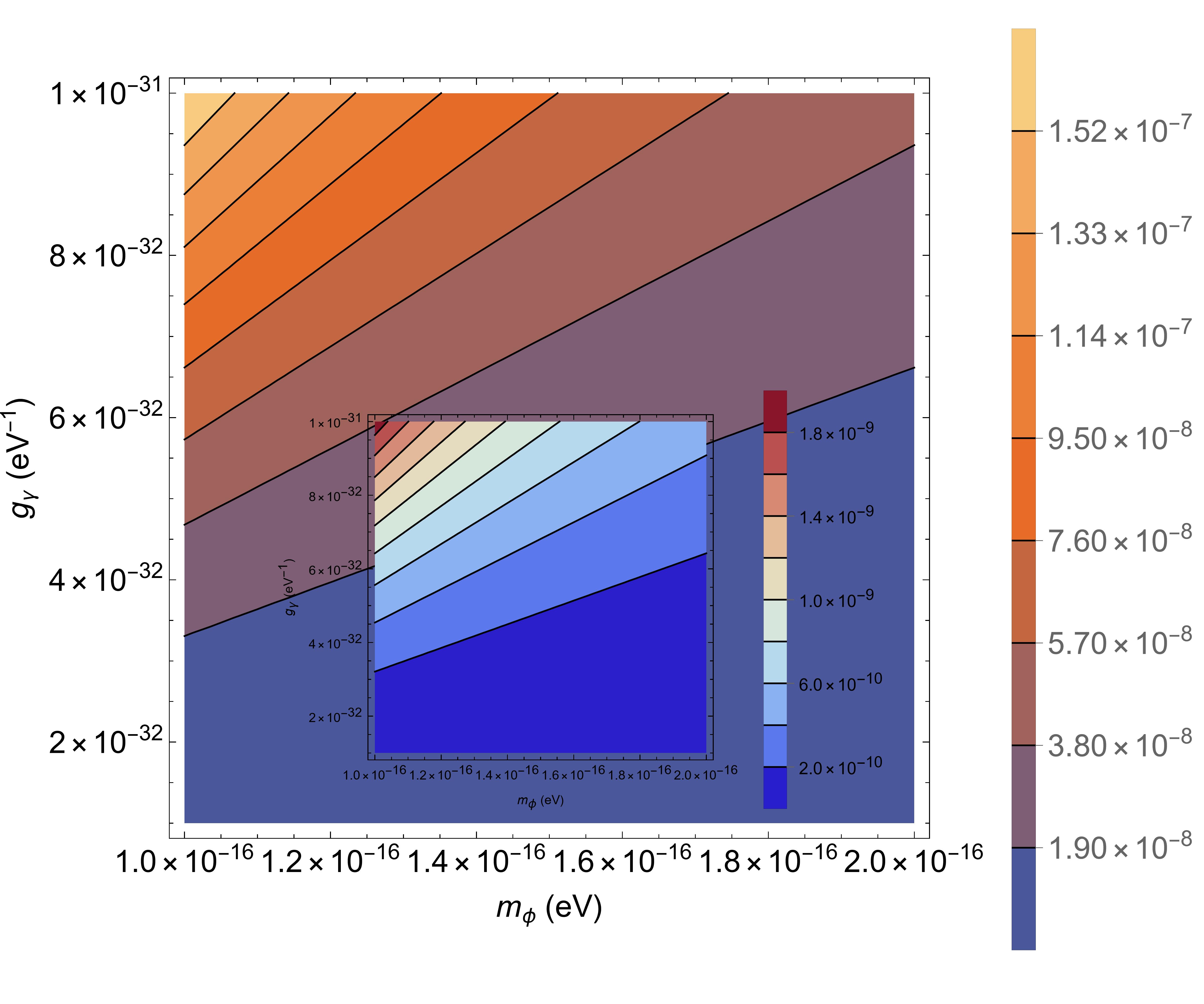}
 \caption{(Color online). Plot of $ \frac{\Delta N}{N_{in}}$ as function of $g_\gamma\in [10^{-32}, 10^{-31}] \, \mathrm{eV^{-1}}$, and $m_\phi\in [  10^{-16}, 2 \cdot 10^{-16}] \, \mathrm{eV}$. Main picture: we consider the squeezing parameter $r=1.7$, which corresponds to $15\; dB$ squeezed states of light. Inset picture:  we use   $r=0.6$, which is compatible with the value  of the squeezing parameter of LIGO experiment. We utilized the same values of $L$ and $\lambda$ considered in Fig. \ref{fig2}.}
\label{fig3}
\end{figure}

\begin{figure}[H]
 \includegraphics[width=\linewidth]{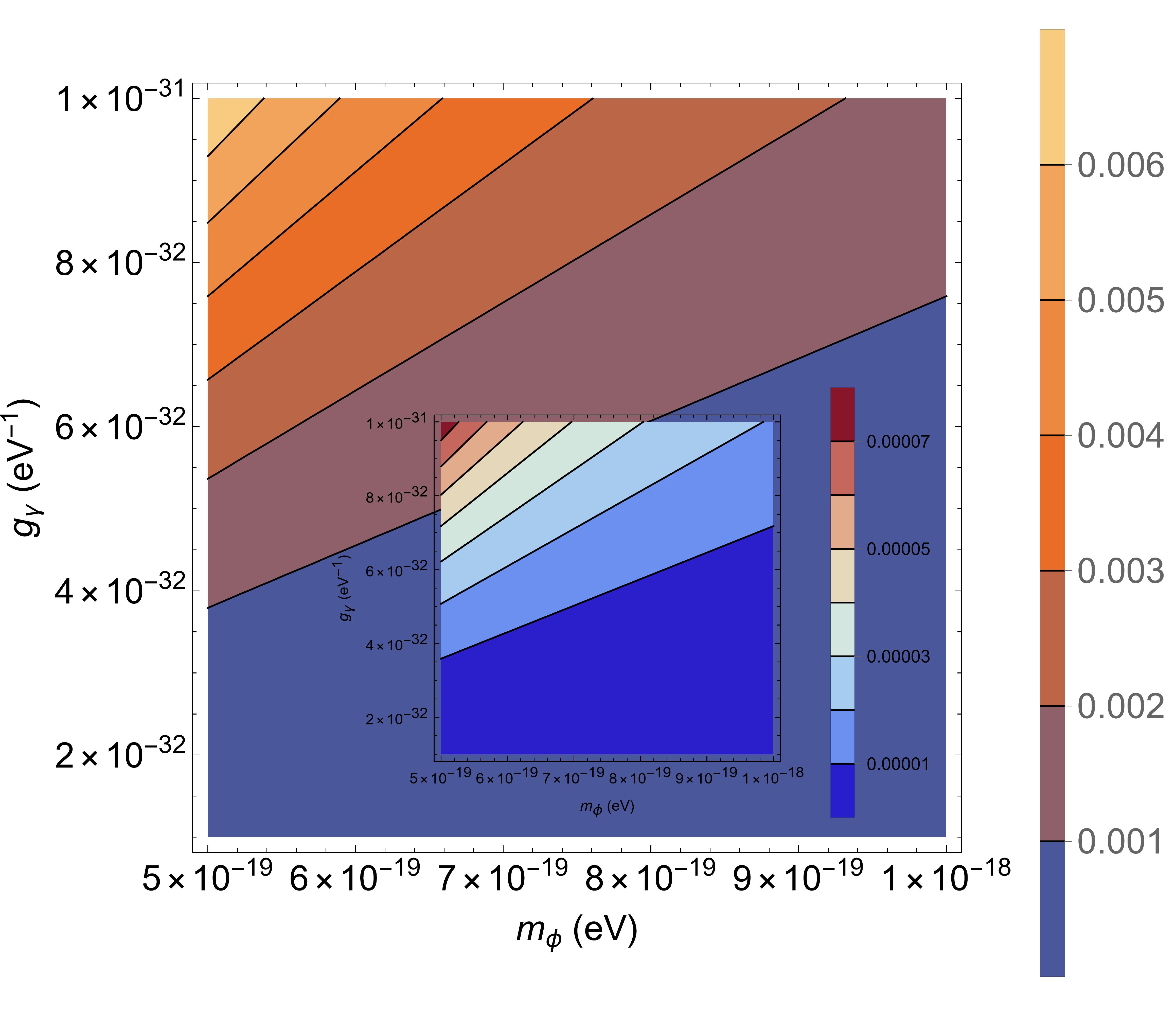}
 \caption{(Color online). Plot of $ \frac{\Delta N}{N_{in}}$ as function of $g_\gamma\in [10^{-32}, 10^{-31}] \, \mathrm{eV^{-1}}$, and $m_\phi\in [0.5\cdot 10^{-18}, 10^{-18}] \, \mathrm{eV}$. Main picture: we consider  $r=1.7$. Inset picture:    $r=0.6$. The same values of $L$ and $\lambda$ considered in Figs. \ref{fig2} and \ref{fig3} are used. }
    \label{fig4}
\end{figure}

\begin{figure}[H]
 \includegraphics[width=\linewidth]{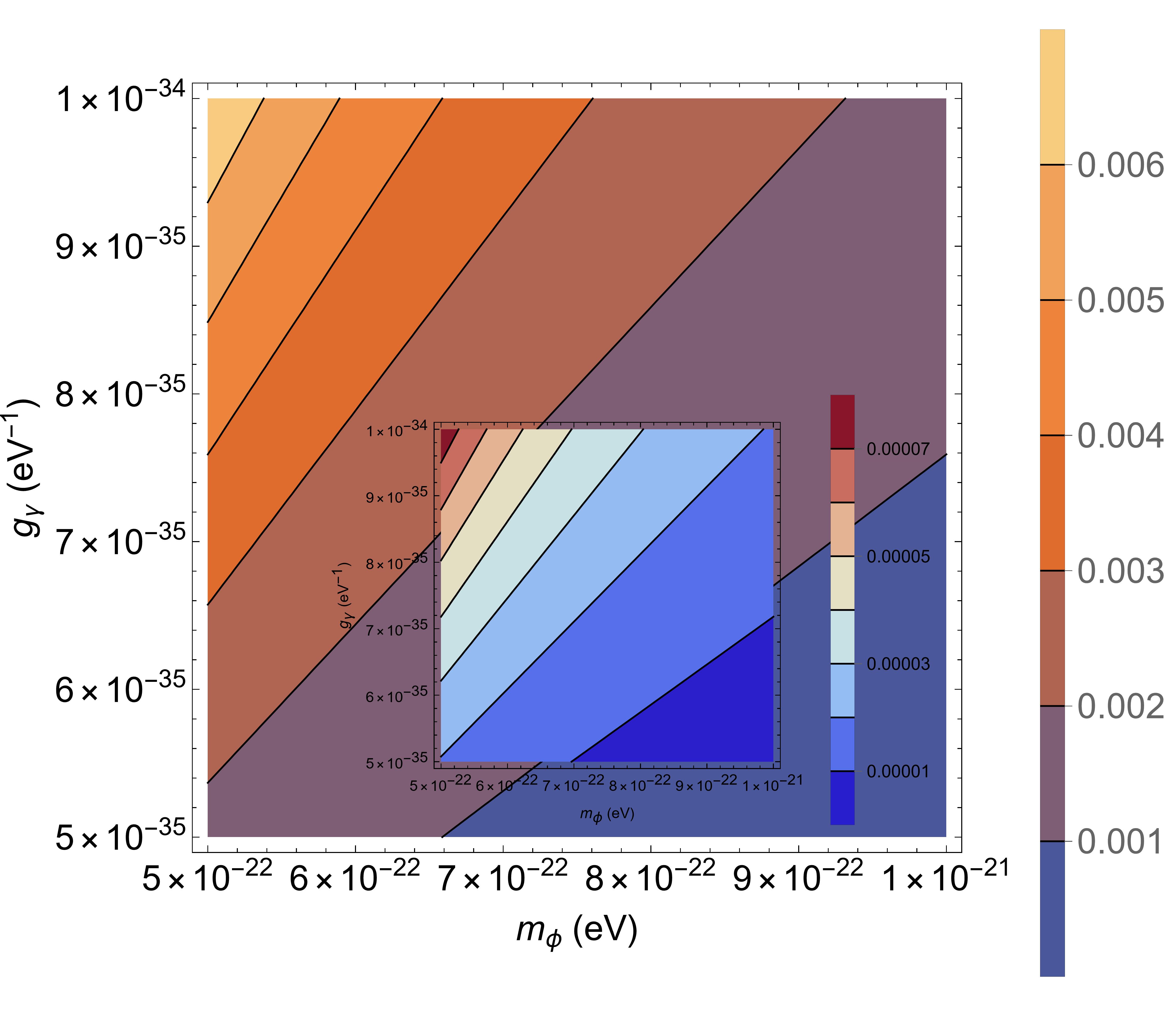}
 \caption{(Color online). Plot of $ \frac{\Delta N}{N_{in}}$ as function of $g_\gamma\in [5 \cdot 10^{-35}, 10^{-34}] \, \mathrm{eV^{-1}}$, and $m_\phi\in [ 5\cdot 10^{-22}, 10^{-21}] \, \mathrm{eV}$. These values of the masses are typical of fuzzy cold dark matter \cite{Fuzzy}.  Main picture: we consider  $r=1.7$. Inset picture:    $r=0.6$. The same values of $L$ and $\lambda$ considered in Figs. \ref{fig2} and \ref{fig3} are used. }
    \label{fig5}
\end{figure}

\begin{figure}[H]
    \centering
    \includegraphics[width=1\linewidth]{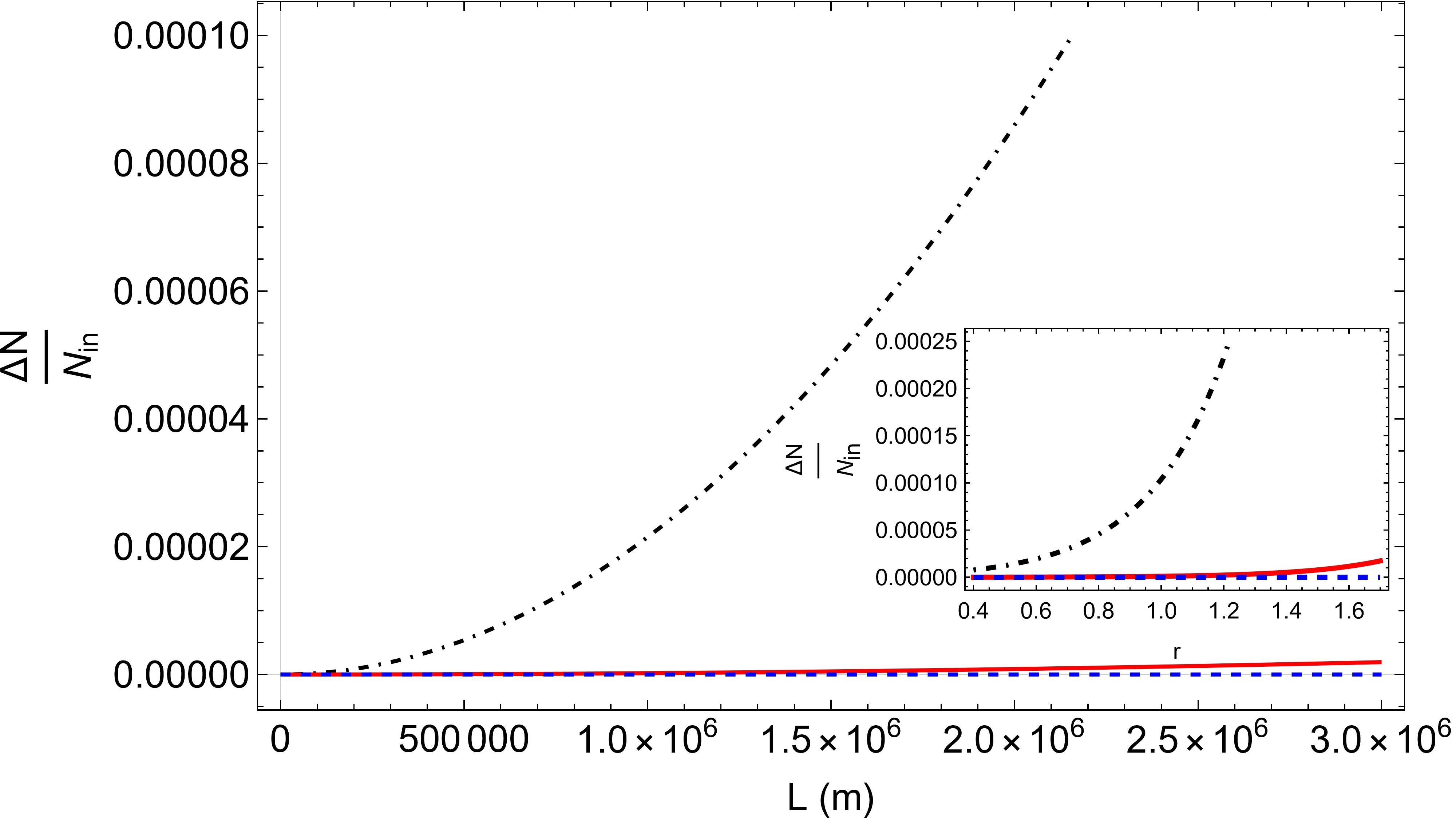}
    \caption{(Color online). Main picture: plot of $ \frac{\Delta N}{N_{in}}$ as function of the arm lenght $L$, for $g_\gamma = 10^{-33} \, \mathrm{eV^{-1}}$ and mass $m_\phi =10^{-20}\; \mathrm{eV}$ (the black dotted line), for $g_\gamma =10^{-32} \, \mathrm{eV^{-1}}$ and mass $m_\phi =10^{-18}\; \mathrm{eV}$ (the red line) and for $g_\gamma = 10^{-30} \, \mathrm{eV^{-1}}$ and mass $m_\phi =10^{-12}\; \mathrm{eV}$ (the blue dashed line), for value of squeezing parameters $r=0.6$.  In the inset is plotted $ \frac{\Delta N}{N_{in}}$  as function of $r$ for $L=300\;\mathrm{km}$ and for
    the  values of $g_\gamma$ and $m_\phi$ used in the corresponding line of the same colors.}
    \label{fig6}
\end{figure}

In our analysis, we consider $\rho_{\text{local}} = 4\cdot10^5 \mathrm{GeV}/m^3$ and the following range of values of $g_\gamma$ and  mass $m_\phi$   \cite{ED4,ED5}:
$
    g_\gamma \in [10^{-35}, 10^{-28}] \, \mathrm{eV^{-1}},$ $m_\phi \in [10^{-22}, 10^{-12}] \, \mathrm{eV}.
$
\\
In Fig. \ref{fig2},  we plot $\delta\left(L\right)$  as function of $m_\phi$ and $g_\gamma$, respectively. We consider    $L = 950 \, \mathrm{km}$, which is compatible with the effective length  of the LIGO interferometer \footnote{The effective lenght is approximatly computed as $L=\frac{2 \mathcal{F}}{\pi}\, 4\mathrm{km}$, where $\mathcal{F}$ is the cavity finesse. This approximation breaks down when the frequency of SDM exceed $1/\tau$ where $\tau$ is the cavity storage time. Considered that storage time of the order $10 \,ms$ the approximation breaks down for $\omega_\phi\geq 50\,Hz$. All the masses considered in this work lie below $10^{-13}\, \mathrm{eV}$ corrisponding to $\approx24\,Hz$. Therefore the effective lenght employeed here is compatible with mass range considered. } \cite{LIGO}. This length is also used in the other figures. A bigger effective length can be achived by using deviced like PVLAS, in which can be reached an effective lenght of $1.47 \cdot 10^9 \, \mathrm{m} $ \cite{PVLAS}. In principle this device can be used for our purposes simply by introducing squeezer-anti squeezer operators at the beginning and at the end of the effective optical path (We point out that in the case of PVLAS the presence of the magnetic field is not necessary in our configuration).
In Figs. \ref{fig3}, \ref{fig4}  and \ref{fig5}, we plot the quantity $\Delta N(L) / N_{in}$, where $\Delta N(L) = N_{out}(L)-N_{in}$ and
\begin{equation}\label{AAA}
    \frac{\Delta N(L)}{N_{in}} =  \sin^2{\delta(L)}\left[\left(\frac{1}{|\alpha|^2}+2 \right)\sinh^2{2r}+ \sinh{4r} \right],
\end{equation}
as  function of   $m_\phi$ and $g_\gamma$ for fixed values of the length $L$ and of the squeezing parameter $r$. The signal we are interested in is the $\Delta N(L)$ which quantifies the impact of scalar interaction on the output photon number. Assuming a Poissonian photon counting statistics, the corrisponding noise is $\sqrt{N_{out}}$ then the signal-to-noise ratio (SNR) is given by
\begin{align}\label{SNR}
\nonumber \text{SNR} &= \frac{\Delta N(L)}{\sqrt{N_{out}}}\\
&=\frac{|\alpha|^2 \sin^2{\delta(L)}\left[\left(\frac{1}{|\alpha|^2}+2 \right)\sinh^2{2r}+ \sinh{4r} \right]}{\sqrt{N_{out}}}.
\end{align}
We point out that $\Delta N(L)$ is zero for $\delta$ and $r$ equal to zero, since in this case $N_{out}(L)=N_{in}$.
We consider $r=0.6$ which is compatible with the value  of the squeezing parameter of LIGO experiment and $r=1.7$, which corresponds to $15 \;dB$ squeezed states of light \cite{INT5}. In Fig. \ref{fig6},   $\Delta N(L) /N_{in}$ is plotted as a function of   $L$, for fixed values of $m_\phi$, $g_\gamma$ and $r$. In the inset of Fig. \ref{fig6}, $\Delta N(L) /N_{in}$ is shown as a function of   $r$, for fixed values of $m_\phi$, $g_\gamma$ and $L$.
In LIGO experiments, the initial number of photons is $N_{in}\sim 10^{23}$  \cite{LIGO} and the noise is approximately $\sqrt{N_{out}}$. In the ranges of the parameters $g_{\gamma}$ and  $m_{\phi}$ analyzed,   $N_{out}$ is of the same order of $N_{in}$ and $\Delta N > \sqrt{N_{out}}$. Another source of uncertainty is represented by the optical losses, which increases the quantum noise in detectors like LIGO and PVLAS. How to reduce this optical losses has been analyzed in \cite{ED6}, where the noise associated to such reduction is estimated in 42 and 87 parts per million, in any case less than $\Delta N/N_{in}$. Other form of noise as thermal noise and seismic noise are expected to be subleading with respect to poissonian noise $\sqrt{N_{out}}$. The instrument sensitivity varies significantly depending on the device considered (LIGO, VIRGO, PVLAS, etc.). Therefore, our plots show that  with gravitational wave detectors, and PVLAS like detectors, the interaction between SDM  (including the fuzzy cold dark matter \cite{Fuzzy}) and photons can be revealed, eventually constraining  the parameters of the SDM model.


\begin{figure}[H]
	\centering
	\includegraphics[width=1\linewidth]{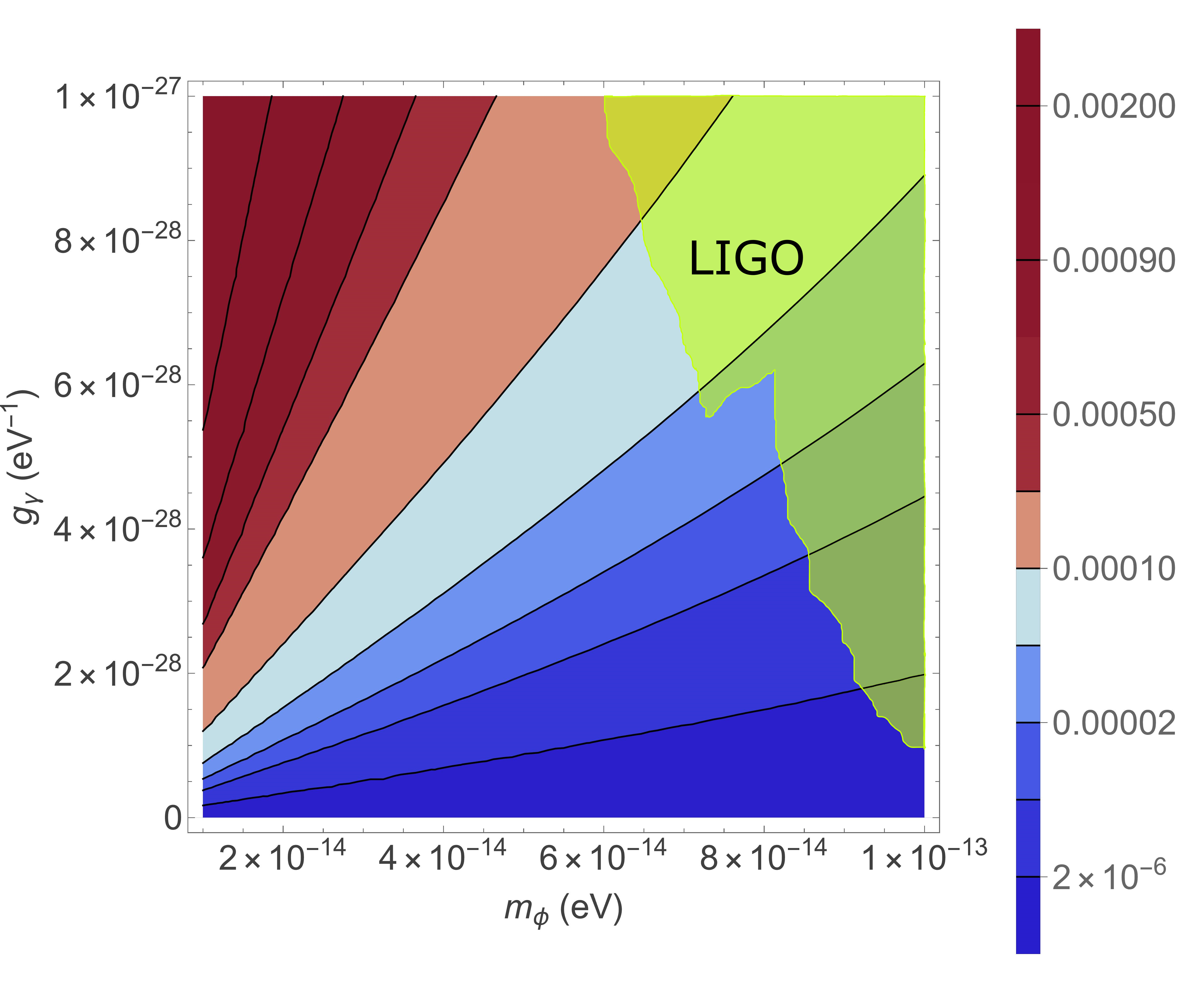}
	\caption{(Color online). Plot of $ \frac{\Delta N}{N_{in}}$ and comparison with the results of \cite{ED7} (green shaded region)  as function of $g_\gamma\in [10^{-33}, 10^{-27}] \, \mathrm{eV^{-1}}$, and $m_\phi\in [ 10^{-14}, 10^{-13}] \, \mathrm{eV}$.}
	\label{fig7}
\end{figure}

In Fig. \ref{fig7} we compare the plots obtained with our method with the one derived in the recent studies regarding upper limits on SDM as derived from LIGO \cite{ED7}, and compare our bindings to their prediction. As it is evidente from Fig. \ref{fig7} the approach we propose can probe a region of parameters space encompassing the range considered in \cite{ED7} and beyond.

\section{ Discussions and conclusions}
 The single-arm interferometric scheme, here introduced,   provides a novel strategy to probe scalar dark matter (SDM) through its coupling to photons. Limitations of the present analysis deserve careful consideration.

First, the scalar dark matter field has been modeled  as a coherently oscillating, monochromatic, and spatially homogeneous classical field [see Eq.(\ref{phase})]. This approximation is valid only within the coherence time and coherence length associated with the galactic dark matter velocity distribution. In realistic scenarios, the finite velocity dispersion of SDM induces a frequency spread and phase decoherence, which may reduce the effective phase accumulation over the interferometer arm length
$ L$. Since the predicted phase shift in Eq. (\ref{eq:phase_shift}) scales linearly with the interaction time (or equivalently with
$L$), such effects could modify the signal profiles shown in Figs. 2–6, especially for the longest effective optical paths considered.

Second, the analysis of the interferometric signal presented  relies on an idealized description of the experimental setup. In particular, the expressions for the output photon number [Eq. (\ref{Numm})] and for the relative variation
$ \Delta N /N_{in}
$,  [Eq. (\ref{AAA})] neglect several realistic noise sources, including optical losses, residual phase noise, technical noise associated with squeezing generation, and fluctuations in the squeezing angle. While we argue   that Poissonian photon-counting noise is subdominant in the parameter regions explored in Figs. 3–6, a complete noise budget analysis is required to quantitatively assess the experimental sensitivity.

Third, the proposed detection strategy critically depends on the implementation of two spatially separated squeezing and anti-squeezing operations, as illustrated schematically in Fig. 1. In realistic interferometers, finite squeezing levels, imperfect mode matching, and optical losses can degrade the effective signal encoded in $\Delta N$. Since the enhancement of the signal scales strongly with the squeezing parameter
$
r$ [see Eqs. (\ref{Numm})–(\ref{SNR}) and Fig. 6], deviations from the ideal Bogoliubov transformations in Eq.(4) may lead to a   reduction of the observable effect.

Moreover, the interaction   described by the effective coupling term
$\phi F^{\mu\nu}F_{\mu\nu}/\Lambda_{\gamma}$, is generic to scalar fields coupled to electromagnetism and is not unique to cosmological dark matter. As a consequence, the signal characterized by the phase shift in Eq.(6) and by the photon-number variation shown in Figs. 3–5 may, in principle, also arise from other scalar backgrounds or from effective variations of electromagnetic parameters. Additional discriminants—such as characteristic temporal modulations linked to Earth’s motion through the galactic halo—will therefore be necessary to unambiguously attribute a detected signal to scalar dark matter.

Despite these limitations, the results presented, demonstrate that single-arm interferometry combined with nonclassical states of light can probe regions of the SDM parameter space that are complementary to those explored by existing interferometric searches. Future work will focus on extending the present framework by incorporating realistic noise models, accounting for the stochastic nature of the SDM field, and developing a concrete experimental implementation within existing or planned interferometric facilities. In this perspective, the approach proposed here may represent a promising and complementary avenue for probing ultra-light scalar dark matter beyond current experimental techniques.

In conclusion,
we presented a novel approach, based on single arm interferometry, to prove the existence of scalar dark matter. Specifically, we investigated the possible coupling between scalar dark matter and photons. We have shown that, should this interaction be present, a single-arm interferometer  would detect a phase shift and an intensity variation, directly resulting from the scalar-photon coupling. The numerical analysis of the phase shift and of the intensity variation indicates that the proposed approach can probe SDM  for a wide range of parameters.
Our setup can be developed by using gravitational wave detectors and two spatially separated squeezing operations, and it
 can in principle be employed to   test  any interaction proportional to the photon number operator.

\appendix

\section{Interaction Hamiltonian}
\label{app:derivation}

In this appendix, we derive the interaction Hamiltonian density in eq.(\ref{Ham}), starting from the Lagrangian density $\mathcal{L}_{\text{int}} =\frac{1}{4} \frac{\phi}{\Lambda_\gamma} F_{\mu \nu} F^{\mu \nu}$. We consider radiation gauge \cite{ItzyksonZuber}, and we set  $\eta_{\mu \nu}=diag(1,-1,-1,-1)$.

Notice that $\mathcal{L}_{\text{int}}=-\frac{\phi(t)}{\Lambda_\gamma} \mathcal{L}_0$,
where $\mathcal{L}_0= -\frac{1}{4} F_{\mu \nu} F^{\mu \nu}=\frac{1}{2}\left(\vec{E}^2-\vec{B}^2\right)$,
is the free electromagnetic Lagrangian.

 As a consequence, $\mathcal{H}_{\text {int }}=-\frac{\phi(t)}{\Lambda_\gamma} \mathcal{H}_0$, where $\mathcal{H}_0=\frac{1}{2}\left(\vec{E}^2+\vec{B}^2\right)$, is the free electromagnetic Hamiltonian.

We derive explicitly this result. We consider  the radiation gauge  $\vec{\nabla} \cdot \vec{A} = 0$ and $A^0 = 0$, where $\vec{A}$  is the vector potential defined by $\vec{B} = \vec{\nabla} \times \vec{A}$, then the electric field is given by $\vec{E} = -\dot{\vec{A}}$. The canonical momentum conjugate to the generalized coordinate $\vec{A}$ is:
\begin{equation}
	\Pi^\mu = \frac{\partial \mathcal{L} }{\partial \dot{A}_\mu} = - \left(1-\frac{\phi(t)}{\Lambda_\gamma}\right) F^{ 0 \mu},
	\label{eq:momentum}
\end{equation}
or
\begin{equation}
	\Pi^0 = 0\quad\text{and}\quad \Pi^i= \left(1-\frac{\phi(t)}{\Lambda_\gamma}\right)E^i,
\end{equation}
where $i=1,2,3$ denote the spatial component. The Hamiltonian density is $\mathcal{H}  = \Pi_\mu  \dot{A}^\mu - \mathcal{L} $.
Being $\dot{\vec{A}} = -\vec{E}$ and using Eq. \eqref{eq:momentum}, one has:
\begin{align}
	\mathcal{H}  =\frac{1}{2}\left( 1-\frac{\phi(t)}{ \Lambda_\gamma}\right) (\vec{E}^2 + \vec{B}^2).
	\label{eq:H_physical}
\end{align}
 The quantum field operators $\hat{\vec{E}}$ and $\hat{\vec{B}}$ are given respectively by \cite{ItzyksonZuber}
\begin{small}
\begin{align}
	\nonumber \hat{\vec{E}}(\vec{r},t) &= \int \frac{d^3k}{\sqrt{2|\vec{k}|(2\pi)^3}} \sum_{\lambda} i|\vec{k}|\bigg[ \hat{a}_{\vec{k}\lambda} \vec{\epsilon}_{\vec{k}\lambda} e^{-ikx}
	 - \hat{a}^\dagger_{\vec{k}\lambda} \vec{\epsilon}_{\vec{k}\lambda} e^{ikx} \bigg], \\
	\hat{\vec{B}}(\vec{r},t) &= \int \frac{d^3k}{\sqrt{2|\vec{k}|(2\pi)^3}} \sum_{\lambda} i|\vec{k}| \bigg[ \hat{a}_{\vec{k}\lambda}(\hat{k} \times \vec{\epsilon}_{\vec{k}\lambda}) e^{-ikx} \nonumber \\
	&\quad - \hat{a}^\dagger_{\vec{k}\lambda} (\hat{k} \times \vec{\epsilon}_{\vec{k}\lambda}) e^{ikx} \bigg],
	\label{field}
\end{align}
\end{small}where $\vec{\epsilon}_{\vec{k}\lambda}$ are the polarization  vectors, and $\hat{a}_{\vec{k}\lambda}$ are the annihilator satisfying the commutation relations $[\hat{a}_{\vec{k}\lambda}, \hat{a}^\dagger_{\vec{k}'\lambda'}] =\delta^{(3)}(\vec{k} - \vec{k}') \delta_{\lambda\lambda'}$. Introducing eqs.\eqref{field} in eq.\eqref{eq:H_physical} and integrating on the volume, we obtain:
\begin{equation}\label{aaa}
	\hat{H}(t) =\hat{H}_{0}+\hat{H}_{int} = \left(1-\frac{\phi(t)}{\Lambda_\gamma}\right) \int d^3k \sum_{\lambda} |\vec{k}| \hat{a}^\dagger_{\vec{k}\lambda}\hat{a}_{\vec{k}\lambda},
\end{equation}
where the second term in eq.(\ref{aaa}) is the interaction Hamiltonian in eq.\eqref{Ham}.

We point out that, in quantum mechanics, if no mixing occurs between annihilation and creation operators, as in the Bogoliubov transformation (eq.(4)), the scalar-photon interaction, eq. \eqref{Ham}, does not produce any observable effect. Indeed, when the squeezing-antisqueezing operation is removed, the expected number of output photons coincides with the number of input photons: see eq. (5), in which $N_{out}=N_{in}$, for $r=0$ and $\delta=0$, which correspond to the absence of Bogoliubov transformations. While in principle also classic techniques may be employed, they must necessarily involve a Bogoliubov transformation and a form of squeezing as in eq. (4). The efficiency of such techniques in highlighting the effect of the SDM is, in each case, dependent on an effective squeezing parameter $r$. We note that our $\text{SNR}$, eq. (8), has no strict classical correspondence, since in our case, $\text{SNR}=0$, for $r=0$ and $\delta=0$. In conclusion, our approach relies on quantum rather than classical aspects.

\section*{Acknowledgements}
A.C., A.Q. and R.S. acknowledge partial financial support from MIUR and INFN. A.C. also acknowledge the COST Action CA1511 Cosmology and Astrophysics Network for Theoretical Advances and Training Actions (CANTATA).


\begin{thebibliography}{99}

\bibitem{DM1}
V.C. Rubin and W.K. Ford Jr., Astrophysical Journal, {\bf 159}, 379 (1970).

\bibitem{DM2}
V.C. Rubin, W.K. Ford Jr., N. Thonnard, Astrophysical Journal, {\bf 238}, 471-487 (1980).

\bibitem{DM3}
V. Trimble, Annual Review of Astronomy and Astrophysics, {\bf 25}, 425-472 (1987).

\bibitem{DM4}
E. Corbelli and P. Salucci, Monthly Notices of the Royal Astronomical Society, {\bf 311}, 441-447 (2000).


\bibitem{DM5}
A. Suarez,  V. H. Robles and T. Matos, Astrophysics and Space Science Proceedings, {\bf 38}, 107-142, (2013).

\bibitem{DM6}

S. Clesse and J. Garcia-Bellido, Phys. Dark Univ., {\bf 22},  137-146 (2018).

\bibitem{DM7}
S. Tulin, and H. B. Yu, Phys. Rep., {\bf 730}, 1-57 (2018).

\bibitem{DM8}
T. Dawoodbhoy,  P. R. Shapiro and  T. Rindler-Daller, Monthly Notices of the Royal Astronomical Society, {\bf 506}, 2418–2444 (2021).




\bibitem{AX1}
 R.D. Peccei, H. Quinn, Phys. Rev. Lett., {\bf 38}, 1440 (1977).

 \bibitem{AX2}
 R.D. Peccei, H. Quinn, Phys. Rev. D, {\bf 16}, 1791 (1977).

 \bibitem{AX3}
  S. Weinberg, Phys. Rev. Lett., {\bf 40}, 223 (1978).

 \bibitem{AX4}
 F. Wilczek, Phys. Rev. Lett., {\bf 40}, 279 (1978).

\bibitem{AX5}
J. E. Kim, Phys. Rev. Lett., {\bf 43}, 103 (1979).


\bibitem{AX6}
M. Shifman, A. Vainshtein and V. Zakharov, Nucl. Phys. B, {\bf 166}, 493 (1980).


\bibitem{AX7}
A. R. Zhitnitsky, Sov. J. Nucl. Phys.,
{\bf 31}, 260 (1980).

\bibitem{AX8}
M. Dine, W. Fischler, and M. Srednicki, Phys.
Lett. B, {\bf 104}, 199 (1981).

 \bibitem{AX9}
G. G. Raffelt, J. Phys. A, {\bf 40}, 6607 (2007).

 \bibitem{AX10}
D. J. E. Marsch, Phys. Rep., {\bf 643}, 1-79 (2016).


 \bibitem{AX11}
A. Capolupo, G. Lambiase, A. Quaranta and S. M. Giampaolo,
 Phys. Lett. B, {\bf 804}, 135407 (2020).





\bibitem{AXE1}
C. Hagmann, P. Sikivie, N. Sullivan, D. B. Tanner and
 S. Cho, Rev. Sci. Instrum., {\bf 61}, 1076 (1990).

\bibitem{AXE2}
C. Hagmann, P. Sikivie, N. S. Sullivan and D. B. Tanner, Phys. Rev. D, {\bf 42}, 1297 (1990).

\bibitem{AXE3}
R. Bradley, J. Clarke, D. Kinion, L. J. Rosenberg, K. van
 Bibber, S. Matsuki, M. Muck and P. Sikivie, Rev. Mod. Phys., {\bf 75}, 777 (2003).


\bibitem{AXE4}
K. Zioutas et al. (CAST), Phys. Rev. Lett., {\bf 94},
121301 (2005).


\bibitem{AXE5}
S. Andriamonje et al. (CAST), JCAP, {\bf 04}, 010 (2007).


\bibitem{AXE6}
S. J. Asztalos et al., Phys. Rev. Lett., {\bf 104}, 041301
 (2010).

\bibitem{AXE7}
J. Hoskins et al., Phys. Rev. D, {\bf 84}, 121302 (2011).


\bibitem{AXE8}
B. T. McAllister, G. Flower, E. N. Ivanov, M. Goryachev, J. Bourhill and M. E. Tobar, Phys. Dark Univ., {\bf 18}, 67 (2017).

\bibitem{AXE9}
V. Anastassopoulos et al. (CAST),
Nature Phys., {\bf 13}, 584
(2017).

\bibitem{AXE10}
T. Braine et al. (ADMX Collaboration), Phys. Rev. Lett., {\bf 124}, 101303 (2020).

\bibitem{AXE11}
C. Bartram et al. (ADMX Collaboration), Phys. Rev.
 D, {\bf 103}, 032002 (2021).

\bibitem{AXE12}
 R. Khatiwada et al. (ADMX), Rev. Sci. Instrum.,
 {\bf 92}, 124502 (2021).

\bibitem{AXE13}
C. Bartram et al. (ADMX Collaboration), Phys. Rev. Lett., {\bf 127}, 261803 (2021).



\bibitem{AXE14}
A. Quiskamp, B. T. McAllister, P. Altin, E. N. Ivanov,
 M. Goryachev and M. E. Tobar, Science Adv., {\bf 8}, 3765 (2022).








\bibitem{susy1}
J. L. Lopez, K. Yuan and D.V. Nanopoulos, 
Nucl. Phys. B, {\bf 267}, 2 (1991).


\bibitem{susy2}
P. Nath and R. Arnowitt,
Phys. Rev. D, {\bf 56}, 5 (1997).


\bibitem{susy3}
S. P. Martin, 
JHEP, {\bf 18}, 1 (1998).


\bibitem{susy4}
J. L. Feng, K. T. Matcvhev, F. Wilczek, 
Phys. Lett. B, {\bf 482}, 4 (2000).

\bibitem{susy5}
S. M. Carroll, 
Liv. Rev. in Relativity, {\bf 4}, 1 (2001).


\bibitem{susy6}
M. E. Gomez, G. Lazarides, C. Pallis, 
Nucl. Phys. B, {\bf 638}, 165-185 (2002).


\bibitem{susy7}
K. Enqvist, A.Mazumdar, 
Phys. Rep., {\bf 380}, 99-234 (2003).

\bibitem{susy8}
T. Cohen, M. Lisanti, A. Pierce, T. R. Slatyer, 
 JCAP, {\bf 2013}, 061 (2013).

\bibitem{susy9}
J. Fan, M. Reece, 
JHEP, {\bf 2013}, 124 (2013).


\bibitem{susy10}
P. Huang, C. E. M. Wagner, 
Phys. Rev. D, {\bf 90}, 015018 (2014).

\bibitem{susy11}
M. Low, L.T. Wang, 
JHEP, {\bf 2014}, 161 (2014).

 \bibitem{susy12}
H. Baer, V. Barger, P. Huang, X. Tata, 
JHEP, {\bf 2012}, 109 (2012).


\bibitem{susy13}
M.J. Dolan, F. Kahlhoefer, C. McCabe et al., 
JHEP, {\bf 2015}, 171 (2015).


\bibitem{susy14}
C. Munoz, 
EPJ web of Conferences, {\bf 36}, 01002 (2017).


\bibitem{susy15}
L. Roszkowski, E. M. Sessolo, S. Trojanowski, 
Rep. Prog. Phys., {\bf 81}, 066201 (2018).



\bibitem{susy16}
 M. Van Beekveld, S. Caron, R. R. De Austri,
  JHEP, {\bf 2020}, 147, (2020).

\bibitem{susy17}
A. Capolupo, G. Pisacane, A. Quaranta and F. Romeo, Phys. Dark Universe, \textbf{46}, 101688 (2024).

\bibitem{DM10}
S.M. Vermeulen,   P. Relton, H. Grote, et al.,
Nature {\bf 600}, 424, (2021).


\bibitem{SDM1}
C. Boehm, P. Fayet, Nucl. Phys. B, {\bf 683}, 219-263 (2004)



\bibitem{SDM2}
X.G. He, T. Li, X. Q. li, J. Tandean, H.-C. Tsai, Phys. Rev. D, {\bf 79}, 023521 (2009)


\bibitem{SDM3}
T. Hambye, F.S. Ling, L.L. Honorez, J. Rocher, JHEP, {\bf 07}, 090, (2009)



\bibitem{SDM4}
M. Kadastik, K. Kannike, M. Raidal, Phys. Rev. D, {\bf 81}, 015002 (2010)

\bibitem{SDM5}
W.L. Guo, Y.L. Wu, JHEP, {\bf 2010}, 83 (2010)

\bibitem{SDM6}
J.M. Cline, P. Scott, K. Kainulainen, C. Weniger, Phys. Rev. D, {\bf 92}, 039906 (2013)


\bibitem{SDM7}
M.R. Buckley, D. Feld, D. Goncalves, Phys. Rev. D, {\bf 91}, 015017 (2015)



\bibitem{SDM8}
K.V. Tilburg, N. Leefer, L. Bougas, D. Budker, Phys. Rev. Lett., {\bf 115}, 011802 (2015)


\bibitem{SDM9}
W. Rodejohann, C. E. Yaguna,  JCAP {\bf 12}, 032, (2015)



\bibitem{SDM11}
J.W. Kee, EPJ Web of Conf., {\bf 168}, 06005 (2018)



\bibitem{Fuzzy}
L. Hui, J. P.  Ostriker, S. Tremaine,  E. Witten,   Phys. Rev. D, {\bf 95}, 043541, (2017).

L. Hui,  Annu. Rev. Astron. Astrophys., {\bf 59}, 247, (2021).



\bibitem{BSM1}
 A. Capolupo,
  Adv.  High Energy Phys., {\bf 2016}, 8089142,   (2016)

A. Capolupo, S. Carloni, A. Quaranta, Phys. Rev. D, {\bf 105},   105013 (2022).

\bibitem{BSM2}
A. Capolupo, A. Quaranta, Phys. Lett. B, {\bf 840}, 137889 (2023).

\bibitem{BSM3}
A. Capolupo, A. Quaranta, R. Serao, Symmetry, {\bf 15}, 807, (2023).


\bibitem{BSM4}
A. Capolupo, A. Quaranta, Phys. Lett. B \textbf{839}, 137776 (2023).


\bibitem{BSM5}
A. Capolupo, G. De Maria, S. Monda, A. Quaranta, R. Serao, Universe, {\bf 10}, 170 (2024).


\bibitem{BSM6}
A. Capolupo, A. Quaranta, J. Phys. G., \textbf{51}, 105202 (2024).

\bibitem{BSM7}
A. Capolupo, S. Capozziello, G. Pisacane, A. Quaranta, Phys. Dark Universe, {\bf 47}, 101748 (2025).




\bibitem{OP1}
E. Savalle, A. Hees, F. Frank, E. Cantin, P.E. Pottie,
 B. M. Roberts, L. Cros, B. T. McAllister, and P. Wolf, Phys. Rev. Lett., {\bf 126}, 051301 (2021).

\bibitem{GW1}
 A. Branca et al., Phys. Rev. Lett., {\bf 118}, 021302 (2017).

\bibitem{GW2}
 H. Grote and Y. V., Phys. Rev. Research, {\bf 1}, 033187 (2019).

\bibitem{GW3}
 S. M. Vermeulen et al., Nature phys., {\bf 600}, 424 (2021).


\bibitem{AE1}
Y. V. Stadnik and V. V. Flambaum, Phys. Rev. Lett., {\bf 115}, 201301 (2015).

\bibitem{AE2}
A. Caputo, G. Ra elt, and E. Vitagliano, Phys. Rev. D, {\bf 105}, 035022 (2022).


\bibitem{DAMA1}
 R. Bernabei et al., Int. J. Mod. Phys. A, {\bf 21}, 1445 (2006).

\bibitem{DAMA2}
H. B. Tran Tan, A. Derevianko, V. A. Dzuba, and V. V.  Flambaum, Phys. Rev. Lett., {\bf 127}, 081301 (2021).

\bibitem{ATOM1}
D. Antypas, O. Tretiak, A. Garcon, R. Ozeri, G. Perez,
 and D. Budker, Phys. Rev. Lett., {\bf 123}, 141102 (2019).

\bibitem{ATOM2}
D. Antypas et al., Quantum Sci. Technol., {\bf 6}, 034001 (2021).

\bibitem{ATOM3}
 S. Aharony, N. Akerman, R. Ozeri, G. Perez, I. Savoray,
 and R. Shaniv, Phys. Rev. D, {\bf 103}, 075017 (2021).

\bibitem{ATOM4}
R. Oswald et al., Phys. Rev. Lett., {\bf 129}, 031302 (2022).

\bibitem{ATOM5}
O. Tretiak et al., Phys. Rev. Lett., {\bf 129}, 031301 (2022).

\bibitem{RING}
A. Ringwald, Phys. of the Dark Universe,  {\bf 1}, 116-135(2012).
\bibitem{ED7}
A. S. G\"{o}ttel et al., Phys. Rev. Lett., {\bf 133}, 101001 (2024).

 \bibitem{INT1}
B. Yurke, S.L. McCall, J.R. Klauder, Phys. Rev. A, {\bf 33}, 4033 (1986).




\bibitem{ED2}
B.P. Abbott et al., [LIGO scientific and VIRGO], Phys. Rev. Lett., {\bf 116}, 061102 (2016).




\bibitem{INT2}
F. Hudelist, J. Kong, C. Liu, et al., Nature Commun., {\bf 5}, 3049 (2014).

\bibitem{INT3}

H. Vahlbruch, M. Mehmet, K. Danzmann and R. Schnabel, Phys. Rev. Lett., {\bf 117}, 110801 (2016).

\bibitem{INT4}
Z. Y. J. Ou, Quantum Optics for Experimentalists, World Scientific (2017)


\bibitem{ED4}
S. Vermeulen et al., Nature phys., {\bf 600}, 424 (2021).



\bibitem{Doli?ska}
A. Doliska et al., Phys. Rev. A, {\bf 68}, 052308  (2003).

\bibitem{mc}
Kirk McKenzie et al., J. Opt. B: Quantum Semiclass. Opt., {\bf 7}, S421 (2005).

\bibitem{Simon}
S. Chelkowski et al., Phys. Rev. A, {\bf 71}, 013806 (2005).

\bibitem{PVLAS}
F. Della Valle et al., EPJ C, {\bf 76}, 24 (2016).


\bibitem{LIGO}
C. L. Mueller, et al.
Rev.  Scient. Instr.,  {\bf 87}, 014502 (2016).

\bibitem{INT5}
X. Xiao et al., Commun. Theor. Phys., {\bf 71}, 037 (2019).


\bibitem{INT6}

A. Buikema et al., Phys. Rev. Lett., {\bf 102}, 062003 (2020).

\bibitem{ED4}
L. Aiello, J. W. Richardson, S. M. Vermeulen, H. Grote, C. Hogan, O. Kwon, and C. Stoughton, Phys. Rev. Lett., {\bf 128}, 121101 (2022).

\bibitem{ED5}
K. Fukusumi, S. Morisaki, and T. Suyama, Phys. Rev. D, {\bf 108}, 095054 (2023).

\bibitem{ED6}
Y. Zhao et al., Phys. Rev. Applied, {\bf 22}, 054040 (2024).


\bibitem{SS7}
W. J. G. de Blok, Advances in Astronomy, {\bf 2010}, 789293 (2010).

\bibitem{SS8}
M. Boylan-Kolchin, J. S. Bullock, and M. Kaplinghat, Monthly Notices of the Royal Astronomical Society, {\bf 415}, 40-44, (2011).

\bibitem{ItzyksonZuber}
C. Itzykson and J.-B. Zuber, {\it Quantum Field Theory} (McGraw-Hill, New York, 1980).

\end{thebibliography}
\end{document}